\documentclass[12pt]{article}
\usepackage{epsf}
\usepackage{graphicx}

\newcommand{\be}{\begin{equation}}
\newcommand{\ee}{\end{equation}}
\newcommand{\bea}{\begin{eqnarray}}
\newcommand{\eea}{\end{eqnarray}}
\newcommand{\beq}{\begin{equation}}
\newcommand{\eeq}{\end{equation}}
\newcommand{\ba}{\begin{array}}
\newcommand{\ea}{\end{array}}
\newcommand{\beqa}{\begin{eqnarray}}
\newcommand{\eeqa}{\end{eqnarray}}

\newcommand{\cL}{{\cal L}}
\newcommand{\cA}{{\cal A}}
\newcommand{\cH}{{\cal H}}
\newcommand{\cO}{{\cal O}}
\newcommand{\cN}{{\cal N}}
\newcommand{\cC}{{\cal C}}

\newcommand{\no}{\nonumber}
\newcommand{\lsim}{\stackrel{<}{_\sim}}

\newcommand{\veps}{\varepsilon}
\newcommand{\MH}{\Lambda_H}
\newcommand{\rH}{\frac{M_\rho^2}{\Lambda_H^2} }
\renewcommand{\Re}{{\rm Re}}
\renewcommand{\Im}{{\rm Im}}

\def\npb#1#2#3{    {\it Nucl. Phys. }{\bf B #1} (#2) #3}
\def\plb#1#2#3{    {\it Phys. Lett. }{\bf B #1} (#2) #3}

\def\prd#1#2#3{    {\it Phys. Rev. }{\bf D #1} (#2) #3}

\def\prl#1#2#3{    {\it Phys. Rev. Lett. }{\bf #1} (#2) #3}

\def\rmp#1#2#3{    {\it Rev. Mod. Phys. }{\bf #1} (#2) #3}

\def\jhep#1#2#3{   {\it JHEP  }{\bf #1} (#2) #3}

\begin{document}

\thispagestyle{empty}
\begin{flushright}
October 2003
\end{flushright}
\vskip 2.5 true cm 

\begin{center}
{\Large\bf On the short-distance constraints from $K_{L,S} \to \mu^+\mu^-$} 
 \\ [25 pt]
 {\sc Gino Isidori}${}^{1)}$ and {\sc Ren\'e Unterdorfer}${}^{1,2)}$
 \\ [20 pt]
${}^{1)}${\sl INFN, Laboratori Nazionali di Frascati, I-00044 Frascati, Italy} 
 \\ [5 pt]
${}^{2)}${\sl Institut f\"ur Theoretische Physik, Universit\"at Wien, 
A--1090 Wien, Austria }
 \\ [25pt]
{\bf Abstract} 
\end{center}
\noindent
Motivated by new precise results on several $K_L$ decays, 
sensitive to the ${K_L \to \gamma \gamma}$ form factor, we present a new 
analysis of the $K_L\to \mu^+\mu^-$ long-distance amplitude
based on the semi-phenomenological approach of Ref.~\cite{DIP}.
Particular attention is devoted to the evaluation of the uncertainties of this method 
and to the comparison with alternative approaches. Our main result is a conservative 
upper bound of $2.5 \times 10^{-9}$ on $B(K_L \to \mu^+ \mu^-)_{\rm short}$, which is  
compatible with the SM expectation and which provides significant constraints on 
new-physics scenarios. The possibility to extract an independent short-distance 
information from future searches on $K_S \to \mu^+ \mu^-$
is also briefly discussed.

\def\thefootnote{\arabic{footnote}}
\setcounter{footnote}{0}
\setcounter{page}{0}

\bigskip
\section{Introduction}
The rare decays $K_{L,S} \to \mu^+ \mu^-$ are a very 
useful source of information on the short-distance 
structure of $\Delta S=1$ flavor-changing neutral-current (FCNC)
transitions. In both cases 
the decay amplitude is not dominated by the clean 
short-distance contribution; however, long-
and short-distance components are comparable in size. 
As a result, even with a limited knowledge of the 
long-distance component, it is possible to extract 
significant constraints on the short-distance part. 

At present the $K_L \to \mu^+ \mu^-$ decay is particularly
interesting in this perspective because of the very precise 
determination of its decay rate \cite{E871}. Here 
the key issue to extract bounds on the short-distance amplitude  
is the theoretical control of the  $K_L \to \gamma\gamma$
form factor with off-shell photons. An effective strategy 
to reach this goal is the combination of phenomenological
constraints from $K_L \to \gamma \gamma$,  $K_L \to \gamma \ell^+\ell^-$ 
and $K_L \to \mu^+\mu^+ e^+e^-$ data, with theoretical 
constraints from chiral symmetry, large $N_C$, and 
perturbative QCD, proposed in Ref.~\cite{DIP}. As far as the phenomenological
constraints are concerned, the situation has substantially 
improved in the last few years thanks to precise results  
on all the $K_L$ Dalitz modes \cite{KTeV_mm,KTeV_mmee,KTeV_ee}
and, to a minor extent, because of the improvements 
on $K_L \to \gamma\gamma$ \cite{NA48,KLOE}.
On the theory side, alternative and/or complementary 
approaches about the high-energy constraints on the 
$K_L \to \gamma\gamma$ form factor have been discussed 
in Ref.~\cite{KPPR,Greynat,Dumm,Valencia}.

The main purpose of this paper is a new analysis 
of the $K_L \to \mu^+ \mu^-$ two-photon dispersive 
amplitude, taking into account these new experimental 
and theoretical developments. In particular, as far as 
the theoretical constraints are concerned, we shall 
quantify the uncertainty of the method in Ref.~\cite{DIP} 
using a more general parameterization of the form factor. 
Moreover, we shall discuss in detail differences and 
similarities between this approach and the one of 
Ref.~\cite{KPPR,Greynat}. As a result of this 
analysis, we find a conservative 
upper bound of $2.5 \times 10^{-9}$ on $B(K_L \to \mu^+ \mu^-)_{\rm short}$, 
which is compatible with the SM expectation, 
$B(K_L \to \mu^+ \mu^-)^{\rm SM}_{\rm short} \approx 0.9 \times 10^{-9}$,
and which provides significant constraints on new-physics scenarios.

On the experimental side, the situation of the $K_S \to \mu^+ \mu^-$ 
decay is very different with respect to the $K_L \to \mu^+ \mu^-$ one:
this process has not been observed yet, and present experimental
limits are still very far from the SM expectation,
$B(K_S \to \mu^+ \mu^-)^{\rm SM} \approx 5 \times 10^{-12}$ \cite{EP}.
On the theory side, the $K_S \to \mu^+ \mu^-$ amplitude is particularly interesting 
since its short-distance component is  dominated by the CP-violating part 
of the $s \to d \ell^+\ell^-$ amplitude, which is very sensitive to new physics 
and which is poorly constrained so far.  As we shall show, 
future searches of $K_S \to \mu^+ \mu^-$ in the $10^{-11}$ range,
could provide significant constraints on several 
consistent new-physics scenarios. 

The plan of the paper is as follows: in Section 2 we
present the general decomposition of $K_L \to \mu^+ \mu^-$ 
amplitude and branching ratio, and briefly 
review the structure of the short-distance component.
In Section 3 we discuss the evaluation of the 
low-energy coupling $\chi_{\gamma\gamma}$, which
encodes the information on the off-shell
$K_L \to\gamma\gamma$ form factor; using this estimate, 
in Section 4 we analyse the present short-distance 
constraints from $K_L \to \mu^+ \mu^-$. 
Section 5 is devoted to  $K_S \to \mu^+ \mu^-$.
The results are summarized in the Conclusions.

\section{Decomposition of the $K_L \to \mu^+\mu^-$ amplitude}

\subsection{General structure and experimental constraints}

\begin{figure}[t]
    \begin{center}
       \setlength{\unitlength}{1truecm}
       \begin{picture}(4.0,2.5)
       \epsfxsize 4.  true  cm
       \epsfysize 2.5  true cm
       \epsffile{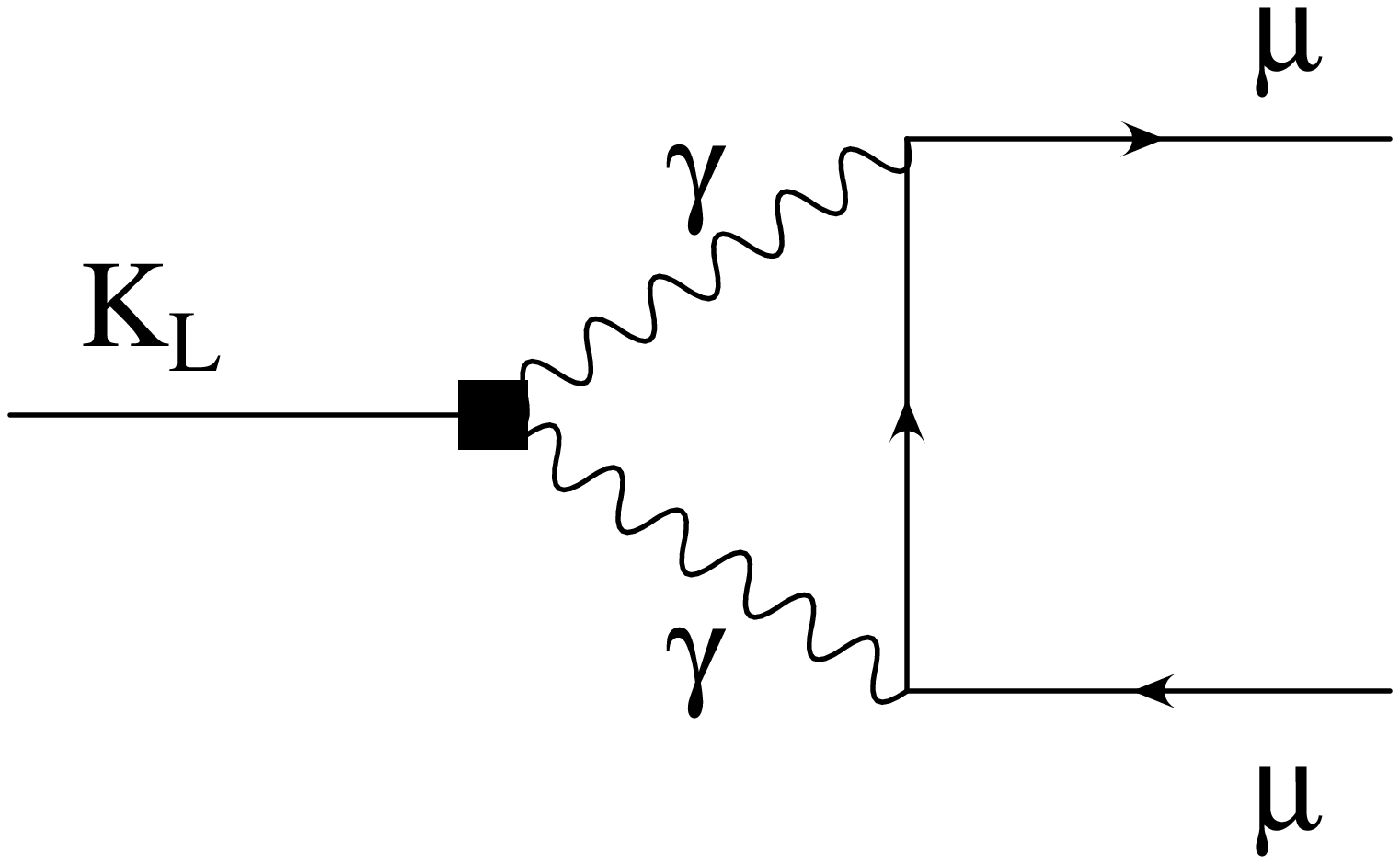}
       \end{picture}
       \centerline{a)}
       \begin{picture}(13.5,2.5)
       \epsfxsize 13.5  true  cm
       \epsfysize 2.5  true cm
       \epsffile{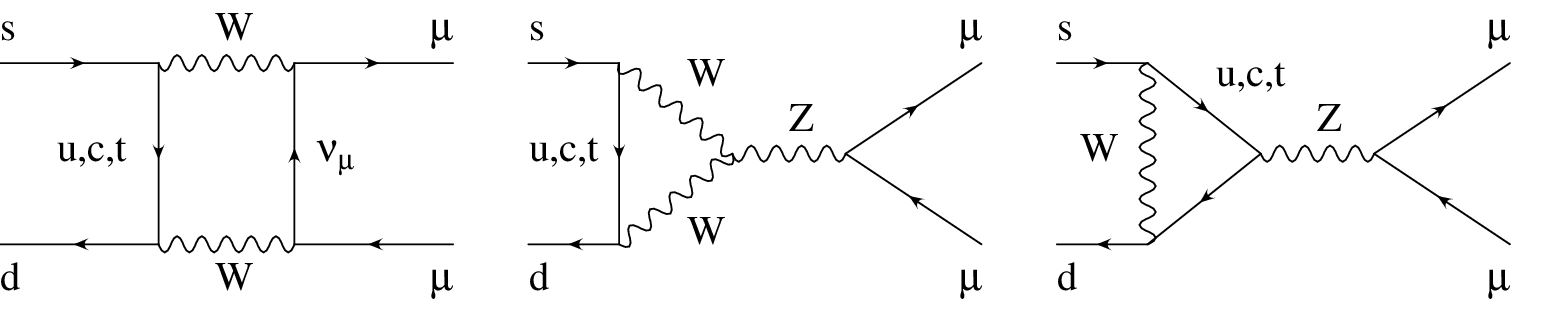}
       \end{picture}       
      \centerline{b)}
    \end{center}
    \caption{Leading long-distance (a) and 
             short-distance (b) contributions to $K_L \to \mu^+\mu^-$.  }
    \protect\label{fig:1}
\end{figure}

Within the Standard Model the $K_L \to \mu^+\mu^-$ decay 
is mediated by a   CP-conserving  
$s$-wave amplitude which can simply be written as
\beq
\cA( K_L \to \mu^+ \mu^- ) = \left( A_{\gamma \gamma} + \Re A_{\rm short} \right)
\bar \mu \gamma_5  \mu~,
\label{eq:one}
\eeq
where $A_{\gamma\gamma}$ denotes the complex long-distance 
contribution generated by the two-photon exchange and 
$\Re A_{\rm short}$ the real amplitude 
of short-distance origin induced by $Z$-penguin and 
box diagrams (see Fig.~\ref{fig:1}).
The long-distance amplitude has a large absorptive part
($\Im A_{\gamma\gamma}$) which is completely dominated by the 
two-photon cut and provides the dominant contribution to 
the total $K_L \to \mu^+\mu^-$ rate.\footnote{~In principle, 
the absorptive amplitude receives additional contributions  
from real intermediate states other than two photons, 
such as two- and three-pion cuts, but these are completely negligible
\protect\cite{deRafael}.}
It is then convenient to normalize $\Gamma(K_L \to \mu^+\mu^-)$
to $\Gamma(K_L \to \gamma\gamma)$ and decompose it as follows
\cite{KPPR,Dumm}
\beqa
\Gamma(K_L \to \mu^+ \mu^-) &=&  \frac{2 \alpha_{\rm em}^2 r_\mu \beta_\mu  }{\pi^2} 
 \left[ R_{\rm abs} + R_{\rm disp} \right] \Gamma(K_L \to \gamma\gamma)~, \label{eq:dec1} \\
R_{\rm abs} &=& (\Im C_{\gamma\gamma})^2 = \left[ \frac{\pi}{2 \beta_\mu }
\ln \frac{1-\beta_\mu}{1+\beta_\mu} \right]^2~, 
 \label{eq:dec2}  \\
R_{\rm disp} &=&  \left[ \chi(\mu) -\frac{5}{2} +\frac{3}{2}\ln\left(\frac{m_\mu^2}{\mu^2}\right)
+ \Re  C_{\gamma\gamma} \right]^2~, 
  \label{eq:dec3} 
\eeqa
where $r_{\mu} = m_\mu^2/m_K^2$, $\beta_{\mu} = \sqrt{ 1-4r_\mu }$ and 
\beq
 C_{\gamma\gamma} = \frac{1}{\beta_\mu} \left[ {\rm Li}_2\left( \frac{ \beta_\mu -1 }{ \beta_\mu+1} \right)
+\frac{\pi^2}{3} +\frac{1}{4} \ln^2 \left( \frac{ \beta_\mu -1 }{ \beta_\mu+1} \right) \right]~.
\eeq
The real coupling $\chi(\mu)=\chi_{\rm short}+\chi_{\gamma\gamma}(\mu)$ encodes 
both short- and long-distance contributions: $\chi_{\rm short}$ denotes
the genuine short-distance (scale-independent) contribution of the diagrams in Fig.~\ref{fig:1}b;
$\chi_{\gamma\gamma}(\mu)$ is a low-energy effective coupling --- determined by the 
behavior of the $K_L \to \gamma\gamma$ form factor outside the physical region --- 
which compensates the scale dependence of the diagram in Fig.~\ref{fig:1}a
(computed with a point-like form factor and regularized in the $\overline{MS}$ scheme).

The smallness of the total dispersive amplitude is well established 
thanks to precise experimental results on both $\Gamma(K_L \to \mu^+ \mu^-)$ 
and $\Gamma(K_L \to \gamma\gamma)$. As pointed out by Littenberg \cite{litt}, 
we can minimize the experimental error on the ratio 
$\Gamma_L(\mu^+ \mu^-)/\Gamma_L(\gamma\gamma)$ 
decomposing it as the product of 
$\Gamma_L(\mu^+ \mu^-)/\Gamma_L(\pi^+\pi^-)=(3.48 \pm 0.05) \times 10^{-6}$ \cite{E871}
times $\Gamma_L(\pi^+\pi^-)/\Gamma_L(\gamma\gamma)$. 
The latter can in turn be determined as follows 
\beq
\frac{  \Gamma_L(\gamma\gamma) }{  \Gamma_L(\pi^+\pi^-) } =
\left\{  \ba{lcl}  \frac{ \Gamma_L(\gamma\gamma) }{ \Gamma_L(\pi^0\pi^0) }
\times \frac{  \Gamma_L(\pi^0\pi^0) }{ \Gamma_L(\pi^+\pi^-) } &=& 0.2798 \pm 0.0044~\cite{litt} \\ 
&& \\
\frac{ \Gamma_L(\gamma\gamma)}{\Gamma_L(3\pi^0) }
\times \frac{ \Gamma_L(3\pi^0)}{\Gamma_L(\pi^+\pi^-) } &=& 0.2836 \pm 0.0060 \ea
\right\} = 0.2811 \pm 0.0035~,
\eeq
where, in the second case, we have used 
$\Gamma_L(\gamma\gamma)/\Gamma_L(3\pi^0)=(2.805 \pm 0.018) \times 10^{-3}$
from the average of NA48 \cite{NA48} and KLOE  \cite{KLOE} 
recent results 
and $\Gamma_L(\pi^+\pi^-)/\Gamma_L(3\pi^0) = (9.89 \pm 0.20)\times 10^{-3}$
from PDG \cite{PDG}. The combination of the two ratios leads to 
\beq
\frac{ \Gamma(K_L \to \mu^+\mu^-) }{ \Gamma(K_L \to \gamma\gamma) }
= ( 1.238 \pm 0.024 )\times 10^{-5}~. \label{eq:R_exp}
\eeq
According to Eqs.~(\ref{eq:dec1})--(\ref{eq:dec3}) this implies 
\beq
R_{\rm disp} = \left[  \chi_{\gamma \gamma}(M_\rho) +\chi_{\rm short}- 5.12 \right]^2  = 0.98 \pm 0.55 \qquad  
{\rm or } \qquad   R_{\rm disp}< 1.7\ {\rm at }\ 90\%\ {\rm CL}~,
\label{eq:bound_R}
\eeq
to be compared with $R_{\rm abs}=27.14$~.

\subsection{The short-distance component}

Within the Standard Model (SM) the short-distance  $K_L \to \mu^+\mu^-$ 
amplitude can be predicted with excellent accuracy in terms of the 
Cabibbo-Kobayashi-Maskawa (CKM) matrix elements $V_{ij}$. 
Following the notation of \cite{BBL}, we find 
\beq
\Re A^{\rm SM}_{\rm short} = - \frac{G_F \alpha_{\rm em}(M_Z) }{ \pi \sin^2 \theta_W }
\sqrt{2} m_\mu F_K \left[ \Re( V_{ts}^* V_{td}) Y(x_t) + \Re(  V_{cs}^* V_{cd} ) Y_{NL} \right]~.
\label{eq:ReAshort} 
\eeq
Employing the decomposition (\ref{eq:dec1})--(\ref{eq:dec3}) this leads to 
\beqa
\chi^{\rm SM}_{\rm short} &=& \kappa \left[ \frac{ - \Re( V_{ts}^* V_{td}) Y(x_t) - \Re(  V_{cs}^* V_{cd} ) Y_{NL}}{4 \times 10^{-4}} \right] = \kappa  (1.11 -0.92 \bar \rho)~,  \label{eq:rho} \\ 
|\kappa| &=& 4\times 10^{-4} \left[ \frac{ m_K}{ 16\pi \Gamma(K_L \to \gamma\gamma) }\right]^{1/2}
 \frac{ \sqrt{2} G_F m_K F_K \alpha_{\rm em}(M_Z) }{ \sin^2 \theta_W  \alpha_{\rm em} } = 1.96~,
\eeqa
where the second identity in (\ref{eq:rho}) has been derived using the 
modified Wolfenstein parameterization of the CKM matrix \cite{CKM} and 
the numerical estimates of $|V_{cb}|$ and $|V_{us}|$ from Ref.~\cite{CKM}
(the two coefficients in Eq.~(\ref{eq:rho}) are affected by a $\sim 5\%$ error,
mainly due to the uncertainty on $|V_{cb}|$ and $Y_{NL}$).
Note that, as explicitly indicated, the value of $\alpha$ in the
short-distance amplitude refers to the electromagnetic coupling 
renormalized at high scales $[\alpha_{\rm em}(M_Z)=1/128]$, 
while in the long-distance part of the amplitude we use 
$\alpha_{\rm em}=1/137.036$~.

Within several SM extensions, the $K_L \to \mu^+\mu^-$ 
amplitude receives additional short-distance contributions
which compete, in magnitude, with the SM one. 
With the exception of rather exotic scenarios,
these new effects induce only a 
redefinition of $\chi_{\rm short}$.
This happens, in particular, in the wide class of models
where the dominant non-standard effects can be encoded 
in effective FCNC couplings of the $Z$ boson: a 
well-motivated framework \cite{CI,BS}
with renewed phenomenological interest \cite{BF}. 
Following the notation of Ref.~\cite{BS}, 
here one can write 
\beq
\chi_{\rm short} \ = \ \chi^{\rm SM}_{\rm short} 
- \kappa \frac{ \Re\left( Z_{ds} -Z_{ds}^{\rm SM} \right)  }{ 4 \times 10^{-4}  }~,
\label{eq:Zds}
\eeq
where $Z_{ds}^{\rm SM}=C_0(x_t) V_{ts}^* V_{td}$ and 
$C_0(x_t) =0.79$.

\section{Theoretical estimate of $\chi_{\gamma\gamma}(\mu)$}

\subsection{The $K_L \to \gamma \gamma$ form factor}
The necessary ingredient to estimate the low-energy coupling 
$\chi_{\gamma\gamma}(\mu)$ is the $K_L \to \gamma\gamma$ 
form factor, $f(q_1^2,q_2^2)$, defined by the four-point 
Green's function
\beq
\int d^4x e^{iq_1 x}
\int d^4y e^{iq_2 y} \langle 0 |T \left\{ J^\mu_{\rm em}(x) J^\nu_{\rm em}(y)
{\cal L}_{\Delta S=1}(0) \right\} | K_L \rangle \propto  \epsilon^{\mu\nu\rho\sigma} 
q_{1\rho} q_{2\sigma}  f(q_1^2,q_2^2)
\label{eq:Green}
\eeq
and normalized such that $f(0,0)=1$. 
In terms of this form factor the  $K_L \to \gamma\gamma$ 
vertex of Fig.~\ref{fig:1} is written as 
\beq
\cA\left[ K_L \to \gamma(\veps_1,q_1) \gamma(\veps_2,q_2) \right] 
= i \frac{ \cN }{ m_K} f(q_1^2,q_2^2)
\epsilon_{\mu\nu\rho\sigma} \veps_1^\mu \veps_2^\nu q_1^\rho q_2^\sigma ~,
\label{eq:Ndef}
\eeq
with $\cN =[ 64\pi\Gamma(K_L\to\gamma\gamma)/m_K]^{1/2} = 1.74 \times 10^{-9}$, and 
\beq
\chi_{\gamma\gamma}(\mu) = \left\{ \frac{ 32 \pi^2 i}{q^2} \int \frac{d^d k}{(4\pi)^d}  
\frac{[ q^2 k^2 -(q \cdot k )^2 ][ f(k^2, (k-q)^2) -1] }{k^2 (k-q)^2[(k-p)^2-m_\mu^2]} 
 \right\}_{\overline{MS}}
\label{eq:chi}
\eeq
with $q^2=m_K^2$ and $p^2=m_\mu^2$. As can be understood from Eq.~(\ref{eq:chi}), 
the leading contribution to $\chi_{\gamma\gamma}(\mu)$ in the chiral expansion
(or in the limit where we neglect external momenta) is governed 
by  $f(-k_E^2, -k_E^2)/k_E^2$, with $k_E^2 \in [0, \infty]$.

\medskip

The structure of $f(q_1^2,q_2^2)$ has been investigated by several authors 
(see e.g. Ref.~\cite{DIP,KPPR,Greynat,Dumm,Valencia,Vari}).
The general properties dictated by QED and QCD can be summarized as follows: 
\begin{enumerate}
\item[i.] it is symmetric under the exchange $q_1^2 \leftrightarrow q_2^2$; 
\item[ii.] it is analytic but for cuts and poles in the region $q_{1,2}^2 > 4m_\pi^2$,
corresponding to the physical thresholds in the photon propagators; 
\item[iii.] the high-energy behavior in the Euclidean region
is given by  $f(-k_E^2, -k_E^2) \sim 1/k_E^2$  (up to logarithmic corrections).
\end{enumerate}
Note that not all the parameterizations proposed in the literature satisfy
these conditions. In particular, the properties ii.~and iii.~are  not 
fulfilled by some of the parameterizations of Ref.~\cite{Valencia}: this 
is one of the main reasons why we do not agree with the 
conclusions of this work.

In addition to these general properties, a systematic and powerful 
approach which leads to tight constraints on the 
structure of  $f(q_1^2,q_2^2)$ is provided by the large $N_C$ expansion~\cite{KPPR}. 
At the lowest order in  $1/N_C$ we expect
\beqa
&& f(q_1^2,q_2^2) = \sum_i \left[  1 + \alpha_i \left( \frac{q_1^2}{q_1^2 - M_i^2} +
\frac{q_2^2}{q_2^2 - M_i^2} \right)  +\beta_i \frac{q_1^2 q_2^2}{ 
(q_1^2 - M_i^2)(q_2^2 - M_i^2)} \right]~,\qquad \label{eq:largeN} \\
&& {\rm with} \quad \sum_i \left[  1 + 2 \alpha_i +\beta_i  \right] =0~, 
\label{eq:largeN2}
\eeqa
where the sum extends over an infinite series of 
(infinitely narrow) vector-meson resonances 
and the sum-rule (\ref{eq:largeN2}) follows 
from the $k_E^2 \to \infty$ condition \cite{KPPR}.
This type of structure is certainly not exact 
point-by-point in $q^2$, but it has been shown 
to provide an excellent tool to evaluate dispersive 
integrals of the type (\ref{eq:chi}), 
even when the sum is truncated to the first non-trivial term 
(see e.g.~Ref.~\cite{KPPR,KPR}). The truncation 
to the lowest vector-meson resonance (the $\rho$ meson)
of Eq.~(\ref{eq:largeN}) corresponds to the DIP ansatz \cite{DIP}
(with $\beta=-1-2\alpha$) and also 
to the choice of Ref.~\cite{Greynat}. Within such 
approximation, the form factor depends on a single 
parameter: in the DIP case this is fixed 
at low energies by the experimental constraints 
on $df(q^2,0)/d q^2$, while in Ref.~\cite{KPPR,Greynat} 
the free parameter is determined by the OPE, matching 
the coefficient of the $1/k_E^2$ term for $k_E^2 \to \infty$.
In the case of pure electromagnetic decays,
such as $\pi^0 \to e^+e^-$ and  $\eta \to \ell^+ \ell^-$, 
these two procedures lead to almost equivalent results.
On the contrary, in the case of  $K_L \to \mu^+ \mu^-$
these two approaches lead 
to a significant numerical difference. 
Once we assume the convergence condition $f(-k_E^2, -k_E^2) \sim 1/k_E^2$, 
the largest contribution to the integral (\ref{eq:chi}) arises 
by the low-energy region. For this reason, we believe that the experimental 
determination of $df(q^2,0)/d q^2$, at low $q^2$, provides the most 
significant constraint on $\chi_{\gamma\gamma}(\mu)$. However, in order to quote
a reliable error on this quantity, we need to estimate also 
the sensitivity of the dispersive integral to the high-energy modes.
To this purpose, in the following we shall employ 
a more general parameterization than the DIP ansatz,
but still inspired by the $1/N_C$ expansion, namely 
\beqa
f(q_1^2,q_2^2) &=&  1 + \alpha\left( \frac{q_1^2}{q_1^2 - M_\rho^2} +
\frac{q_2^2}{q_2^2 - M_\rho^2} \right) - (1+2\alpha-\delta) \frac{q_1^2 q_2^2}{ 
(q_1^2 - M_\rho^2)(q_2^2 - M_\rho^2) } \no\\
&&  + \gamma \left( \frac{q_1^2}{q_1^2 - \MH^2} +
\frac{q_2^2}{q_2^2 - \MH^2} \right) - (2\gamma+\delta) \frac{q_1^2 q_2^2}{ 
(q_1^2 - \MH^2)(q_2^2 - \MH^2) }~,
\label{eq:fpar}
\eeqa
with $\MH >  M_\rho$. The choice of the four coefficients in 
(\ref{eq:fpar}) is such that the sum-rule (\ref{eq:largeN2})
is automatically satisfied and for $\gamma=\delta=0$ we recover 
the DIP case (with $\beta=-1-2\alpha$).

The expression of $\chi_{\gamma\gamma}(\mu)$ obtained by means of the 
parameterization (\ref{eq:fpar}) is
\beqa
&& \chi_{\gamma\gamma}(\mu) = \frac{3}{2} \ln\left(\frac{\mu^2}{M^2_\rho}\right) - R(M^2_\rho,M^2_\rho) 
            +2\alpha\left[R(M^2_\rho,0) -R(M^2_\rho,M^2_\rho)\right] \no \\
        &&\qquad  +\delta\left[R(M^2_\rho,M^2_\rho)-R(\MH^2,\MH^2)\right] 
             +2\gamma\left[R(\MH^2,0)-R(\MH^2,\MH^2)\right]~, \quad
\label{eq:chi_exp}
\eeqa
where, expanding up to the next-to-leading order in the chiral expansion, 
\beqa
&&\!\!\!\!  R(M^2,0) =  -\frac{3}{2} \ln\left(\frac{M^2_\rho}{M^2}\right)
-\frac{3}{2}+\frac{m_K^2}{M^2} \left[ \frac{1-10 r_\mu}{6} \ln\left(\frac{m_\mu^2}{M^2}\right)
- \frac{11+34 r_\mu}{36} \right] +\cO\left( \frac{m_K^4}{M^4} \right)~, \no \\
&&\!\!\!\!  R(M^2,M^2)= -\frac{3}{2} \ln\left(\frac{M^2_\rho}{M^2}\right)
-\frac{m_K^2}{M^2} \frac{1-5r_\mu}{3} +\cO\left( \frac{m_K^4}{M^4} \right)~. 
\eeqa

The result for $ \chi_{\gamma\gamma}(\mu)$ in (\ref{eq:chi_exp}) depends on four free 
parameters: $\alpha$, $\gamma$, $\delta$ and $\MH^2$. 
As anticipated, one combination is fixed by the 
model-independent constraint following by 
the experimental measurement of the form factor 
in the Dalitz decays $K_L \to \gamma \ell^+ \ell^-$
and $K_L \to e^+ e^- \mu^+\mu^-$. 
The KTeV collaboration has performed a detailed 
analysis of this form factor, taking into account 
the significant distortions of the spectrum induced 
by radiative corrections \cite{barker}. As a result, 
a coherent determination of the slope
\beq 
\alpha_{\rm exp} =
 - M^2_\rho \left. \frac{d}{d q^2} f(q^2,0) \right|_{q^2=0}
= \alpha + \rH \gamma
\label{eq:alpha_exp}
\eeq
emerges from all the available channels: 
$K_L \to \gamma e^+ e^-$ ($\alpha_{\rm exp}|_{ee}= - 1.63 \pm 0.05$
\cite{KTeV_ee}), $K_L \to \gamma \mu^+ \mu^-$  
($\alpha_{\rm exp}|_{\mu\mu}= - 1.54 \pm 0.10$ \cite{KTeV_mm}), 
and $K_L \to e^+ e^- \mu^+\mu^-$
($\alpha_{\rm exp}|_{ee\mu\mu} = - 1.59 \pm 0.37$ \cite{KTeV_mmee}).
Combining these results, we obtain the very precise 
value $\alpha_{\rm exp}= - 1.611 \pm 0.044$.
Note that the consistency of the DIP parameterization
for the various modes,
as well as the one of the BMS model \cite{Vari}, 
provides a good support in favor of a generic 
VMD structure, such as the one in Eq.~(\ref{eq:fpar}),   
with a minor role played by the terms associated 
to the heavier poles.

Taking advantage of the relation (\ref{eq:alpha_exp}) 
and expanding up to the first order in the three small ratios
$m_\mu^2/m_K^2$, $m_K^2/M_\rho^2$ and $M_\rho^2/\MH^2$, we can write
\beqa
 \chi_{\gamma\gamma}(M_\rho) &=& - 3 \alpha_{\rm exp} +  \frac{m_K^2}{M^2_\rho}  \left[ 
\frac{1}{3} +\alpha_{\rm exp} \frac{ 1+ 6 \ln (m^2_\mu/M_\rho^2)}{18} \right] 
- \Delta_{\Lambda} = (5.83 \pm 0.15) - \Delta_{\Lambda}~, \quad \no \\
 \Delta_{\Lambda} &=& \left[ 3 \gamma \left(1-\rH\right) 
+ \delta \left(   \frac{m^2_K}{3M_\rho^2} - \frac{3}{2}\ln\rH \right) \right]~. \label{eq:chi_rho2}
\eeqa
From the expression (\ref{eq:chi_rho2}) we deduce that:
i) $\chi_{\gamma\gamma}(M_\rho)$ shows a rapid 
convergence in the chiral expansion;
ii) the combination of  $\gamma$, $\delta$ and $\MH^2$ which 
controls the sensitivity to the high-energy modes is $\Delta_{\Lambda}$.
We also note that  $\Delta_{\Lambda} \ll 1$ 
in the limit $\MH \to M_\rho$ ($\gamma$ and $\delta$ are expected 
to be at most of $\cO(1)$ by naturalness arguments), 
thus the scenario which maximizes the sensitivity to high scales 
is the case $\MH^2 \gg  M^2_\rho$. 
Since charm loops play an important role in determining 
the short-distance behavior of this form factor, we cannot 
exclude a priori that $\MH$ is connected to the charm 
scale. For this reason, in the following we shall 
allow $\MH^2$ to reach values up to $10 M^2_\rho$. 

Estimates of $\Delta_{\Lambda}$ can be obtained 
by looking at the  $k_E^2 \to \infty$ behavior of 
$f(-k_E^2,-k_E^2)$:
\beq
\lim_{k_E^2 \to \infty} f(-k_E^2,-k_E^2) = \frac{2}{k_E^2}
\left[ (1+\alpha-\delta)M_\rho^2 +(\gamma+\delta) \MH^2\right] \doteq
\cC_0 \frac{M_\rho^2}{k_E^2}~.
\label{eq:C0}
\eeq 
If $|\cC_0|\lsim 1$, as it happens in the 
case of the pure electromagnetic form factors relevant to 
$\pi^0(\eta) \to \ell^+ \ell^-$ decays \cite{KPPR}, we must impose 
the condition $\gamma \approx - \delta $. As a first qualitative 
observation, we note that  $\Delta_{\Lambda}$ is rather small
once this condition is fulfilled: $|\Delta_{\Lambda}|<1$ for $\MH^2/M^2_\rho<10$ 
and $|\gamma,\delta|<1$.
A more quantitative estimate along this line can be obtained 
following  Ref.~\cite{Greynat}: bosonizing the partonic currents 
in (\ref{eq:Green}), Greynat and de Rafael argued that 
$\cC_0$ can be computed explicitly 
to leading order in both the $1/N_C$ and the chiral expansion, obtaining
\beq
\cC_0 = \frac{ 8\pi^2 F_\pi^2}{ N_C M^2_\rho} \left[ 1 \pm \cO(L_5) \right] = 0.38 \pm 0.12~
\cite{Greynat}~.
\label{eq:Greynat}
\eeq
Using this constraint in (\ref{eq:C0}) and imposing 
the additional conditions $|\gamma,\delta|<0.5$ and $2< \MH^2/M^2_\rho<10$, 
we find $\Delta_{\Lambda} = 0.3 \pm 0.6$, where the error scales almost 
linearly with the upper bound on $|\gamma,\delta|$~. This result
already provides a good support in favor of the smallness of 
$|\Delta_{\Lambda}|$; however, we believe that an even 
more convincing argument can be obtained by looking 
at the partonic evaluation of $f(q_1^2,q_2^2)$,
which we shall discuss in the following.

\subsection{High-energy behavior of  the 
$K_L \to \gamma \gamma$ form factor from perturbative QCD}

\begin{figure}[t]
    \begin{center}
      \setlength{\unitlength}{1truecm}
       \begin{picture}(13.5,2.5)
       \epsfxsize 13.5  true  cm
       \epsfysize 2.5  true cm
       \epsffile{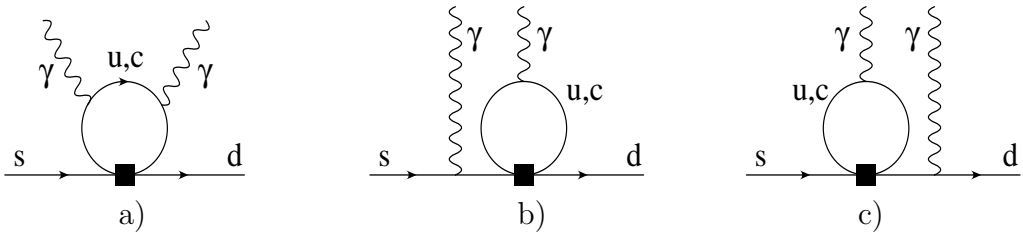}
       \end{picture} \\ 
    \hglue -0.3 true cm a) \hglue 4.8 true cm b) \hglue 4.0 true cm  c) 
    \end{center}
    \caption{Lowest-order partonic diagrams contributing  to the 
    $K_L \to \gamma \gamma$ amplitude. }
    \protect\label{fig:2}
\end{figure}

In the Euclidean region of large $|q^2_{1,2}|$,
with $| q^2_1 -q^2_2 | \gg \Lambda^2_{QCD}$,
the $K_L \to \gamma \gamma$ form factor can 
computed reliably using perturbative QCD. 
Indeed, this kinematical configuration resembles 
the case of a heavy-quark decay into a 
two-body heavy-light system. Here non-perturbative 
effects associated to the hadronization of the 
external quark lines have shown to be factorizable
--- up to power suppressed terms ---
into appropriate hadronic wave functions \cite{BBNS}.
We expect a similar factorization mechanism to hold 
also for the Green's function in (\ref{eq:Green}), 
where the lowest-order partonic kernel corresponds 
to the diagrams in Fig.~\ref{fig:2}, and 
the leading hadronic matrix element 
involved is simply 
\beq
\langle 0 | \bar{s}\gamma_\mu \gamma_5 d | K^0(p) \rangle = i \sqrt{2} F_K p^\mu~.
\eeq

The $K_L \to \gamma\gamma$ amplitude computed in this way,
at the leading-logarithmic level of accuracy, can be written as 
\beqa
\cA\left[ K_L \to \gamma(\veps_1,q_1) \gamma(\veps_2,q_2) \right] 
&=&  i \epsilon_{\mu\nu\rho\sigma} \veps_1^\mu \veps_2^\nu q_1^\rho q_2^\sigma 
\frac{ 16 \alpha_{\rm em} G_F V_{us}^* V_{ud} F_K }{9 \pi \sqrt{2} } \left[C_2(\mu) +3C_1(\mu)\right]
\no \\
&& 
\times \left[ I\left(\frac{q_1^2}{m_c^2},\frac{q_2^2}{m_c^2}\right) 
+ T\left(\frac{q_1^2}{m_c^2}\right) 
+ T\left(\frac{q_2^2}{m_c^2}\right) \right]~,\qquad\quad  
\label{eq:pert}
\eeqa
where $C_{1,2}(\mu)$ are the Wilson coefficients of the $\Delta S=1$ 
effective Hamiltonian, in the notation of Ref.~\cite{BBL}, and 
\beqa
I(r_1,r_2) &=&  \frac{1}{6}+ \int_0^1 dx \int_0^{1-x} dy 
\left\{  \frac{ (2-x-y) + (r_1 x+r_2 y)(1 - x - y)^2 }{ 
                 1 - (r_1 x+r_2 y)(1 - x - y) } \right. \no \\ 
&&  \left.
-(2-3x-3y)\ln\left[ \frac{ (r_1 x+r_2 y)(1 - x - y) - 1 }{ 
  (r_1 x+r_2 y)(1 - x - y) } \right] \right\}~,
\quad  \no \\
T(r) &=& 2  \int_0^1 dx ~
x(1-x) \ln\left[ \frac{  r x(1 -x )-1 }{  r x(1 -x ) } \right]~.
\qquad
\label{eq:IT_nT}
\eeqa
The  Wilson coefficients $C_{1,2}(\mu)$ are understood to 
be computed at the leading-logarithmic level at $\mu \sim (q_1^2+q_2^2)/2$, 
starting from the initial conditions $C_{2}(M_W)=1$ and ${C_{1}(M_W)=0}$.

In the limit $r_{1,2} \to - \infty$ we find
\beqa
I(-r_1,-r_2)     &  \longrightarrow  & 
  \frac{  \ln[2(r_1+r_2)] }{r_1+r_2}
   \left[ 1+ \cO\left(\frac{(r_1-r_2)^2}{(r_1+r_2)^2}\right)\right]~, \no \\ 
T(-r_1)+T(-r_2)  & \longrightarrow &  \frac{1}{r_1+r_2}
   \left[ 1+ \cO\left(\frac{(r_1-r_2)^2}{(r_1+r_2)^2}\right)\right]~,
\label{eq:pert_lim}
\eeqa
which agree with the results obtained in Ref.~\cite{DIP} 
in the limit $q_1^2=q^2_2$. On general grounds, the perturbative
calculation is not reliable in the limit $q_1^2=q_2^2$, since this 
kinematical configuration leads to a soft-momentum kaon. 
This is indeed one of the main criticism 
which could be addressed to the DIP analysis.
However, by means of Eqs.~(\ref{eq:pert_lim}) we can now 
show that the limit $q_1^2=q_2^2$ does not spoil the leading 
$1/q_{1,2}^2$ behavior in the perturbative region 
$|q^2_1 +q^2_2 | \gg | q^2_1 -q^2_2 | \gg \Lambda^2_{QCD}$.
For this reason, we shall use the perturbative result in
(\ref{eq:pert}), in the limit $-q_1^2=-q_2^2=k_E^2 \gg \Lambda^2_{QCD}$, 
to constrain the high-energy behavior of $f(-k_E^2,-k_E^2)$. 

\begin{figure}[t]
\vskip  0.8 cm
$f(-k^2,-k^2)$
\vskip -2.8 cm
$$
\includegraphics[width=15 cm]{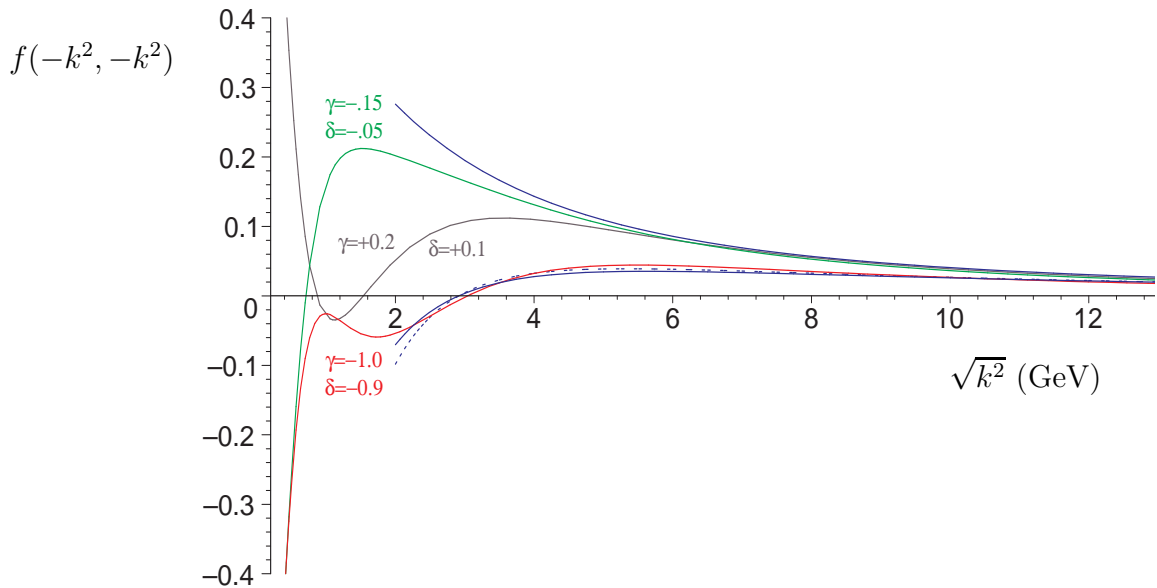}
$$
\vskip  -4.3 cm
\hskip  12.5 cm
$\sqrt{k^2}$~(GeV)
\vskip   3.3 cm
\caption{\label{fig:ff}
Comparison of the perturbative form factor 
[lines starting from $\sqrt{k^2}=2$~GeV: 
upper full curve = no QCD corrections and un-expanded 
loop functions in Eq.~(\ref{eq:IT_nT}); lower full curve = leading-log QCD 
corrections included and un-expanded loop functions;
lower dotted curve = leading-log QCD 
corrections included and expanded loop functions in Eq.~(\ref{eq:pert_lim})] 
with the phenomenological ansatz,
for $\Lambda_H = 3~{\rm GeV}$ and different choices of the parameters (see text).}
\end{figure}

The comparison of  the perturbative form factor with 
the phenomenological parameterization, in the Euclidean region, 
is shown in Fig.~\ref{fig:ff}. As can be noted, with a suitable
choice of parameters it is quite easy to reproduce the 
perturbative behavior, at large $k_E^2$,
with the phenomenological ansatz.  In order to define 
a range of $\gamma$, $\delta$ and $\Lambda_H$ compatible with the 
high-energy  behavior, we have imposed the condition that 
the phenomenological form factor must be in the range defined 
by the perturbative result, with or without QCD corrections,
for $\sqrt{k_E^2} \geq 2$~GeV. This condition leaves 
a considerable uncertainty in the form factor 
in the region  1~GeV~$\lsim \sqrt{k_E^2} \lsim 4$~GeV
(see Fig.~\ref{fig:ff}).
However, the resulting range for the weighted integral 
in (\ref{eq:chi}) is rather limited.
This fact should not surprise, since the kernel 
in (\ref{eq:chi}) enhances the sensitivity to the 
low-energy region. As a result of this condition, we find
\beq
| \Delta_{\Lambda} | \leq 1.0~.
\label{eq:Delta_fin}
\eeq
in good agreement with the conclusions already derived 
starting from Eqs.~(\ref{eq:C0}) and (\ref{eq:Greynat}).

\subsection{Interference of short- and long-distance components}
As can be noted from Fig.~\ref{fig:ff},  
the short-distance constraint on $f(-k_E^2,-k_E^2)$
is not sufficient to determine the sign 
of the physical  $K_L \to \gamma\gamma$
amplitude, or the sign of the coupling $\cN$ in Eq.~(\ref{eq:Ndef}).
By construction, we have assumed $\cN>0$  in the phenomenological analysis, 
hiding this ambiguity in the sign of $\kappa$,
the parameter of Eq.~(\ref{eq:rho})
which controls the interference between 
$\chi_{\gamma\gamma}(\mu)$ and $\chi_{\rm short}$.

Within Chiral Perturbation Theory, the leading contributions
to the on-shell 
$K_L \to \gamma\gamma$ amplitude are the  
tree-level pole diagrams with 
$\pi^0$, $\eta$ and $\eta'$ exchange\footnote{~The $\eta'$ exchange, 
formally of higher order, cannot be 
neglected due to the cancellation of the 
lowest-order $\pi^0$ and $\eta$ diagrams in the limit
where we neglect $\eta$--$\eta'$ mixing and 
apply the Gell-Mann--Okubo mass formula.} \cite{Handbook}.
In all realistic scenarios of pseudoscalar meson mixing, 
the contributions induced by $\eta$ 
and $\eta'$ poles cancel to a large extent
and the $K_L \to \gamma\gamma$ amplitude 
turns out to be dominated by the $\pi^0$ pole
(see e.g.~Ref.~\cite{KLOE,Dumm,Handbook}).
The $\pi^0$ pole contribution to the 
$K_L \to \gamma\gamma$ amplitude can be written as 
\beqa
\cA^{(\pi^0)}( K_L \to \gamma \gamma ) 
= \frac{ i 2 G_8 F_\pi \alpha_{\rm em}  }{\pi} \frac{ m_K^2 }{ m^2_K -m_\pi^2 }
\epsilon_{\mu\nu\rho\sigma} \veps_1^\mu \veps_2^\nu q_1^\rho q_2^\sigma ~,
\label{eq:KL_G8}
\eeqa
where $G_8$ denotes the leading $(8_L,1_R)$ coupling of the 
chiral non-leptonic weak Lagrangian: $\cL_{W}^{(2)} = G_8 {\rm tr}
(\lambda_6 \partial U^\dagger \partial U )$ \cite{Handbook}.
The sign of $G_8$ cannot be determined in a model-independent way; 
however, it can be predicted by the $\Delta S=1$
partonic Lagrangian computing the hadronic matrix 
elements of four-quark operators in the large $N_C$ limit
(or employing naive factorization). In the convention defined 
by Eq.~(\ref{eq:ReAshort}), this procedure 
leads to ${\rm sgn}(G_8) <0$ \cite{Pich2,BDI},
which would imply a {\em negative} interference between 
$\chi_{\gamma\gamma}(\mu)$ and $\chi_{\rm short}$. 
In other words, under the following two assumptions
\begin{description}
\item{i.}  ${\rm sgn}[\cA( K_L \to \gamma\gamma)] = 
            {\rm sgn}[\cA( K_L \to \pi^0 \to \gamma\gamma)] $;
\item{ii.} ${\rm sgn}[\langle \pi^0 | \cH_{W} | K_L \rangle ] = 
  {\rm sgn} [\sum C_i(\mu) \langle \pi^0 | O_i(\mu) | 
K_L \rangle_{N_C \to \infty} ]$; 
\end{description}
we need to set $\kappa <0$. This conclusion agrees with the result 
of Ref.~\cite{Dumm}, where a negative interference 
between $\chi_{\gamma\gamma}(\mu)$ and $\chi^{\rm SM}_{\rm short}$
has also been derived employing large $N_C$ arguments. 
However, it is worth to emphasize that this conclusion depends 
only on the two assumptions stated above, whose validity 
go beyond the $1/N_C$ expansion.

\section{Short-distance constraints from $K_L \to \mu^+\mu^-$}

Taking into account the estimate of $\Delta_{\Lambda}$
in (\ref{eq:Delta_fin}), we can quote as final estimate 
for the low-energy coupling $\chi_{\gamma\gamma}$:
\beq
\chi_{\gamma\gamma}(M_\rho) = 5.83 \pm 0.15_{\rm exp.} \pm 1.0_{\rm th.}
\label{eq:fin_chi_gg}
\eeq
As expected, this result is numerically rather close to 
the DIP one \cite{DIP}, and to the one of Ref.~\cite{Dumm},
while it is substantially different 
from the value $\chi_{\gamma\gamma}(M_\rho) = 2.18 \pm 0.15 \pm 0.9$
quoted in Ref.~\cite{Greynat}. As shown in the previous section, the difference 
between our result and Ref.~\cite{Greynat}
does not arise by the high-energy behavior 
of the $K_L \to \gamma\gamma$ form factor, 
which is essentially the same in both cases,
but it is a consequence of the inclusion 
of the low-energy condition~(\ref{eq:alpha_exp}).

Although the final estimate in (\ref{eq:fin_chi_gg})
is numerically rather similar to the one of Ref.~\cite{DIP}, 
the present analysis is certainly more conservative.
Indeed, the error in (\ref{eq:fin_chi_gg}) is mainly 
of theoretical nature and should be regarded
as the definition of a flat confidence interval,
rather than the standard deviation of a statistical 
distribution. It will be very hard to decrease 
this error only with the help of new experimental 
data on $K_L \to \gamma \ell^+ \ell^-$ and 
$K_L \to e^+ e^- \mu^+ \mu^-$ decays: the only significant 
improvement could arise by the determination 
of the quadratic slope $\beta$ from $K_L \to e^+e^+ \mu^+\mu^-$,
which is beyond the reach of present kaon facilities. 
On the other hand, a significant step forward could in principle 
be obtained by means of Lattice-QCD calculations 
of the form factor in the region  
1~GeV~$\lsim \sqrt{k_E^2} \lsim 4$~GeV.

Using the result (\ref{eq:fin_chi_gg}) in Eq.~(\ref{eq:bound_R}),
and solving the quadratic equation in terms of $\chi_{\rm short}$,
we finally obtain
\beq 
\chi_{\rm short} = \left\{
\ba{r} -1.7 \pm 1.4 \\   0.3 \pm 1.4 \ea  \right.
\qquad \longrightarrow \qquad 
-3.1  <  \chi_{\rm short} <  1.7~.
\label{eq:chi_sh_fin}
\eeq
In the most conservative scenario, i.e. without assumptions about the 
sign of the interference between short and long-distance amplitudes, 
only the upper bound on $|\chi_{\rm short}|$ can be used. From the 
latter we derive the conservative upper bound 
\beq 
 B(K_L \to \mu^+ \mu^-)_{\rm short} <  2.5 \times 10^{-9}~.
\label{eq:bd_chi_short}
\eeq
It is worth to emphasize that 
the result in (\ref{eq:chi_sh_fin}) is perfectly compatible
with the SM also when the sign estimate discussed in the 
previous section is taken into account: 
$\chi^{\rm SM}_{\rm short} =-1.9 \pm 0.2$ (for $\kappa<0$). 
Using the SM expression (\ref{eq:rho}), the bound (\ref{eq:chi_sh_fin})
can be translated into the following range for $\bar \rho$:
\beq
-0.5  < \bar \rho < 2.1\ (2.9)~,
\eeq
where the number between brackets does not take into account 
the sign estimate. 

\begin{figure}[t]
\vskip -0.5 cm 
$$
\includegraphics[height=9 cm]{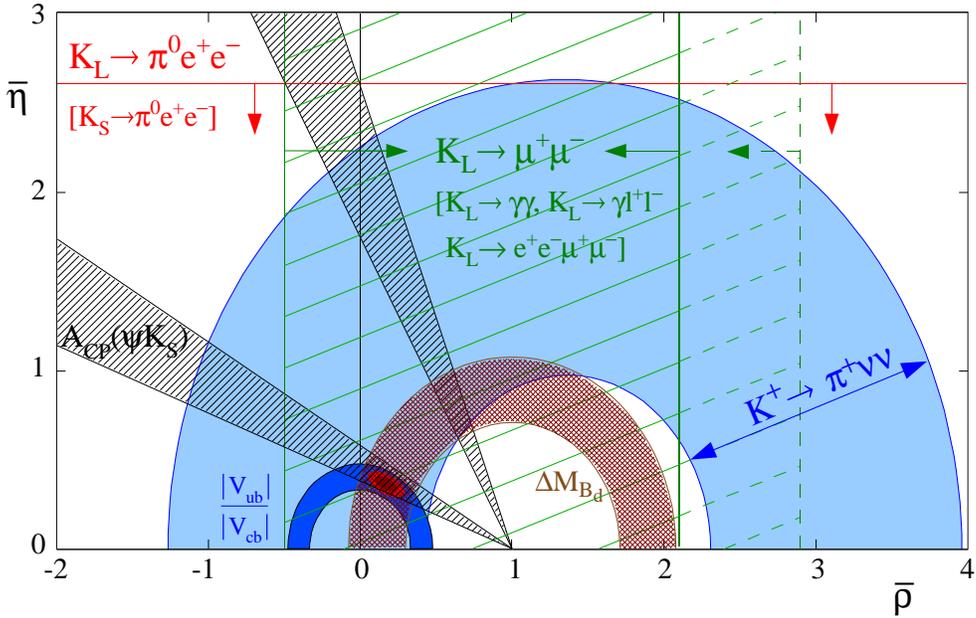}
$$
\vskip -0.5 cm
\caption{ \label{fig:new}
  Summary of the present constraints in the $\bar \rho$--$\bar \eta$
  plane from rare $K$ decays: the $K_L \to \mu^+\mu^-$
  bound on $\bar \rho$ discussed in this work (with and without the  
  sign estimate); the  $K_L \to \pi^0 e^+e^-$ bound 
  on $\bar \eta$~\cite{BDI} and the  $K^+ \to \pi^+ \nu \bar{\nu}$ ellipse~\cite{E787}. 
  The constraints on $|V_{ub}|$ and $B$--$\bar B$ 
  mixing, from Ref.~\cite{CKM}, are also shown for comparison.}
\end{figure}

As can be seen in Fig.~\ref{fig:new}, this bound does not compete 
in precision with present CKM constraints from $B$ physics; however, 
it does compete with constraints from other $\Delta S=1$ FCNC 
processes, such as $K^+ \to \pi^+ \nu \bar{\nu}$. The interest 
of these bounds is their implications for non-standard scenarios. 
For instance, using Eq.~(\ref{eq:Zds}), the  $K_L \to \mu^+\mu^-$ bound 
leads to 
\beq
-2.2\ (-3.3)  <  \frac{ \Re\left( Z_{ds} \right) }{ \Re\left( Z_{ds}^{\rm SM} \right) }  < 2.2~,
\eeq
which is one of the most stringent constraints on possible non-standard
FCNC couplings of the $Z$ boson.

\section{$K_S \to \mu^+\mu^-$}
\label{sect:KS}
The most general decomposition of $K^0~(\bar K^0) \to \ell^+\ell^-$ 
decay amplitudes does not include only the $s$-wave ($A$) component
in Eq.~(\ref{eq:one}), but also a $p$-wave ($B$) term:
\beq
\cA(K^0 \to \ell^+ \ell^-) = \bar u_\ell (iB + A\gamma_5 ) v_\ell~,
\eeq
with a corresponding decay rate given by \cite{Handbook}
\beq
\Gamma(K_{L,S} \to \ell^+\ell^-) = \frac{m_K \beta_\ell}{8\pi}\left(|A|^2
+\beta_\ell^2 |B|^2\right)~, \qquad 
\beta_\ell = \left(1- \frac{4m_\ell^2}{m_K^2}\right)^{1/2}~.
\eeq
The two amplitudes have opposite CP, such that CP-conserving 
contributions to $K_L$ and $K_S$ decays are generated 
by $A$ and $B$, respectively. Long-distance contributions 
generated by intermediate two-photon states 
can lead to both $A$ and $B$ amplitudes, but
in both cases these have a negligible CP-violating component. 
Short-distance contributions 
of the SM type can contribute only to the $A$ 
amplitude, but in this case CP-violating phases 
are expected to be $\cO(1)$. For this reason, within the SM
and in any SM extension with the same basis of 
effective FCNC operators, we can safely  neglect the 
$B$ term in the $K_L\to \mu^+\mu^-$ amplitude
(as we have done so far). 
On the other hand, in the $K_S\to \mu^+\mu^-$
case we need to keep both types of amplitudes 
and we can write
\beq
\Gamma(K_{S} \to \ell^+\ell^-) = \frac{m_K \beta_\ell}{8\pi}
\left[ (\Im A_{\rm short} )^2 + \beta_\ell^2 (\Re B_{\gamma\gamma})^2  + 
\beta_\ell^2 (\Im B_{\gamma\gamma})^2 \right]~.
\label{eq:KS_rate}
\eeq
The remarkable feature of Eq.~(\ref{eq:one}) is the fact that 
the three contributions add incoherently in the total rate. 
It is then much simpler to derive constraints 
on the short-distance component from the experimental 
limit on $\Gamma(K_{S} \to \ell^+\ell^-)$: in the most 
conservative case, we can derive a model-independent 
bound on $|\Im A_{\rm short}|$ simply setting to zero  
the long-distance terms.

Analogously to Eq.~(\ref{eq:ReAshort}), within the SM one finds 
\beq
\Im A_{\rm short}^{\rm SM} = - \frac{G_F \alpha_{\rm em}(M_Z) }{ \pi \sin^2 \theta_W }
\sqrt{2} m_\mu F_K \Im( V_{ts}^* V_{td}) Y(x_t) 
\eeq
which leads to 
\beqa
B(K_{S} \to \mu^+\mu^-)^{\rm SM}_{\rm short} 
&=& 1.0 \times 10^{-5} \times \left| \Im( V_{ts}^* V_{td}) \right|^2 \no \\
&=& 1.4  \times 10^{-12} \times \left| \frac{V_{cb} }{ 0.041}  \right|^4  
\times \left| \frac{\lambda }{ 0.223 }  \right|^2 \times \bar \eta^2 ~. 
\label{eq:KSs}
\eeqa
According to the value of $\bar \eta$ obtained from global CKM fits,  
this contribution is in the $10^{-13}$ range. However, the present constraints 
on $\bar \eta$ derived only from $\Delta S=1$ FCNC processes are rather weak 
(see Fig.~\ref{fig:new}): this implies that, at present, new-physics scenarios 
where $B(K_{S} \to \mu^+\mu^-)_{\rm short}$
reaches the $10^{-11}$ level are perfectly allowed. 

Contrary to the $K_L\to \mu^+\mu^-$ case, the dispersive long-distance $K_S\to \mu^+\mu^-$
amplitude is unambiguously determined at the lowest order in the chiral
expansion: it arises by two-loop diagrams of the type 
$K_S \to \pi^+\pi^- \to \gamma \gamma \to \mu^+\mu^-$, which 
are finite due to the absence of corresponding local terms \cite{EP}.
The dispersive amplitude is about 2.3 times the absorptive one, 
and summing the two contributions Ecker and Pich found \cite{EP}:
\beq
B(K_{S} \to \mu^+\mu^-)_{\rm \gamma \gamma} = 1.9 \times 10^{-6} \times 
B(K_{S} \to \gamma \gamma ) \approx 5 \times 10^{-12}~.
\label{eq:KSl}
\eeq
Due to possible higher-order chiral corrections, 
this prediction is expected to hold within a $30\%$ error.

Comparing Eq.~(\ref{eq:KSs}) and Eq.~(\ref{eq:KSl}), we conclude that 
a search for $K_{S} \to \mu^+\mu^-$ in the $10^{-11}$ range would be 
very useful. An evidence of $K_{S} \to \mu^+\mu^-$ well above 
$10^{-11}$ would be a clear signal of new physics. 
In absence of a signal, as expected within the SM, 
bounds on $B(K_{S} \to \mu^+\mu^-)$
close to $10^{-11}$ could be translated into interesting
model-independent bounds on the CP-violating phase of the $s \to d \ell^+\ell^-$ 
amplitude.\footnote{~These bounds could be very useful to discriminate 
among new-physics scenarios if other modes, 
such as $K^+ \to \pi^+ \nu \bar\nu$, would indicate 
a non-standard enhancement 
of the $s \to d \bar \ell \ell $ transition}

\section{Conclusions}

The short-distance dominated $\Delta S=1$ FCNC transitions
are a key element to investigate the mechanism of quark-flavor 
mixing, and the rare decays $K_{L,S} \to \mu^+ \mu^-$ represent 
one of the most useful experimental probes of these processes.
In this paper we have re-analyzed the extraction of short-distance
constraints from $K_{L} \to \mu^+ \mu^-$. To this purpose,
we have quantified the uncertainty of the long-distance 
$K_{L} \to \gamma\gamma \to \mu^+ \mu^-$ amplitude. The latter 
has been evaluated employing a 
semi-phenomenological approach to the $K_{L} \to \gamma\gamma$
form factor. At short-distances, this approach
is essentially equivalent to the one of Ref.~\cite{KPPR,Greynat};
however, the semi-phenomenological approach 
is superior in the low-energy region, where it takes into 
account the precise experimental constraints on the Dalitz modes. 
As we have shown, this region provides the dominant 
contribution to the dispersive integral needed to 
evaluate the $K_{L} \to \gamma\gamma \to \mu^+ \mu^-$ amplitude.

The short-distance bounds thus derived, expressed as 
effective bounds on the CKM parameter $\bar \rho$, are summarized in 
Fig.~\ref{fig:new}. Although not competitive with $B$-physics
constraints, these bounds provide a serious challenge to many NP models.
The present range is mainly determined by the theoretical error 
of the approach and it will be very hard to decrease it by means of 
experimental data only. A minor improvement could be 
expected with better data on the ratio 
$\Gamma(K_L \to \mu^+ \mu^-)/\Gamma(K_L \to \pi^+\pi^-)$
and thus a smaller error in Eq.~(\ref{eq:bound_R}).
As we have stressed, a major improvement could be 
obtained by means of Lattice-QCD calculations of the $K_{L} \to \gamma\gamma$ 
form factor in the Euclidean  region.

Finally, we have briefly discussed the possibility to 
extract short-distance constrains also from  $K_S \to \mu^+ \mu^-$. 
As we have shown, bounds on $B(K_{S} \to \mu^+\mu^-)$
close to $10^{-11}$ could be translated into interesting
model-independent bounds on the CP-violating phase 
of the $s \to d \ell^+\ell^-$ amplitude.

\newpage

\section*{Acknowledgments}
We thank Gerhard Buchalla for several useful discussions in 
the early stage of this work. We acknowledge also interesting
discussions with Giancarlo D'Ambrosio, Gerhard Ecker, Marc Knecht and 
Eduardo De Rafael. This work is partially supported by IHP-RTN, 
EC contract No.\ HPRN-CT-2002-00311 (EURIDICE).

\bigskip


\end{document}